# Exploring complex graphs using three-dimensional quantum walks of correlated photons

Max Ehrhardt,[1] Robert Keil,[2] Lukas J. Maczewsky,[1] Christoph Dittel,[3] Matthias Heinrich,[1] and Alexander Szameit[1]*.

[1]Institut für Physik, Universität Rostock, Albert-Einstein-Straße 23, 18059 Rostock, Germany

[2]Institut für Experimentalphysik, Universität Innsbruck, Technikerstr. 25, 6020 Innsbruck, Austria

[3]Physikalisches Institut, Albert-Ludwigs-Universität Freiburg, Hermann-Herder-Str. 3, 79104 Freiburg, Germany

*correspondence to: *alexander.szameit@uni-rostock.de*

**Abstract:** Graph representations are a powerful concept for solving complex problems across natural science, as patterns of connectivity can give rise to a multitude of emergent phenomena. Graph-based approaches have proven particularly fruitful in quantum communication and quantum search algorithms in highly branched quantum networks. Here we introduce a new paradigm for the direct experimental realization of excitation dynamics associated with three-dimensional networks by exploiting the hybrid action of spatial and polarization degrees of freedom of photon pairs in complex waveguide circuits with tailored birefringence. This novel testbed for the experimental exploration of multi-particle quantum walks on complex, highly connected graphs paves the way towards exploiting the applicative potential of fermionic dynamics in integrated quantum photonics.

**One Sentence Summary:** Correlated photons undergo quantum walks on 3D graphs governed by their spatial and polarization degrees of freedom.

**Main Text:**

Complex networks occur across many different fields of science, ranging from biological signaling pathways and biochemical molecules exhibiting efficient energy transport (*1*), to neuromorphic circuits and social interactions in the ever-expanding world wide web (*2*). In general, such structures are modelled by graphs, whose complexity is dictated by the number of nodes and linkage patterns between them. Crucially, any physical representation of a graph is constrained by the fact that both of these need to be arranged in three-dimensional space. A dramatic example of the resulting scaling behavior, which is highly unfavorable for physical simulation, is the human brain, in which the already staggering number of 80 billion neurons is dwarfed by the 100 trillion synapses enabling the flow of signals between them (*3*). Despite the comparably miniscule number of nodes, discrete quantum systems face a similar challenge: Complex network topologies with a high degree of connectivity allow for potent applications (*4*) such as efficient multipartite quantum communication (*5*) or search algorithms (*6*, *7*), but the physical implementation of such networks has – so far – been constrained to two dimensions.

The transport properties of connected graphs are usually studied in the framework of quantum walks (QWs). The paradigmatic case of a linear one-dimensional (1D) chain has been implemented on a wide range of different technological platforms (*8–14*). Whereas the dynamics of single-particle QWs can be understood through classical interference phenomena (*15*), QWs of multiple indistinguishable particles (*16*) fundamentally exceed this framework: Their non-classical correlations arising from multi-particle quantum interference (*17*) enable novel applications in quantum computation (*18*), e.g. a solver for the graph isomorphism problem (*19*). Moreover, virtual multidimensional graphs can be obtained by interpreting multi-particle systems as single walkers (*20*, *21*). Utilizing the dimensionality of experimental platforms – supporting the implementation of graphs – as a resource for increasing the vertices' connectivity, sheds light onto coherence phenomena ranging from biological systems (*22–24*) to graph theory problems (*21*), and enables quantum search algorithms (*25*). Simultaneously leveraging both methods – multiple walkers *and* two-dimensional (2D) single-particle graphs – brings the experimental realization of graphs with even higher complexity into reach (*26*). Unfortunately, spatial dimensions are a rather sparse commodity, a fact that is often exacerbated by practical limitations of technological platforms. Internal degrees of freedom of the walkers can mitigate these constraints by providing artificial dimensions (*27*). In the context of (classical wave) photonics, this strategy has enabled the implementation of specific scenarios via coupled sets of physical states (*28*, *29*) or additional parameter dependencies in the system dynamics (*30*). The development of methods that are applicable to multiple walkers, and genuinely increase the dimensionality of any physically implemented underlying graph, remains an open challenge of crucial importance.

In this work, we demonstrate controlled QWs of correlated photons on three-dimensional (3D) graphs. We realize the graph structure by means of a novel hybrid approach: 2D photonic lattices of spatially coupled waveguides are inscribed into fused silica by femtosecond laser writing (*31*), while one synthetic dimension is encoded in the photons' polarization. The dynamics within the synthetic dimension are established by harnessing the intrinsic birefringent properties of the elliptical waveguides. By appropriately orienting their respective fast axes, we realize continuous coupling between the two orthogonal polarization states: A single waveguide yields coherent population oscillations between the polarization modes (Fig. 1**A**), whereas planar waveguide lattices are transformed into bilayer versions of themselves (Fig. 1**B**). Notably, following this strategy to implement higher-dimensional graphs naturally gives rise to hypercube (HC) symmetries that have recently been predicted to leave their distinct signature on the evolution of correlated photon pairs (*32*). In order to discern the fingerprints of quantum interference, we use the nonclassicality $v = C_{\text{ind}}/C_{\text{dis}} - 1$, contrasting the measured coincidence count rates between indistinguishable ($C_{\text{ind}}$) and distinguishable photons ($C_{\text{dis}}$), respectively. We further examine conditions for the selective suppression ($v < 0$) or enhancement ($v > 0$) of output states, revealing bosonic as well as fermionic behavior on certain subgraphs, respectively. Finally, we show that our technique allows for the control of both strength and sign of the coupling along the synthetic dimension, and employ negative couplings to selectively induce partial breaking of the graph's HC symmetry. Our approach opens new avenues for the experimental exploration of quantum dynamics on highly complex graphs that play an essential role in numerous disciplines of science.

In order to illustrate the working principle, we first show that the hallmark of two-particle interference, the Hong-Ou-Mandel (HOM) effect (*33*), arises in the polarization degree of freedom of a single waveguide. Direct laser-written waveguides in fused silica are intrinsically birefringent (*34*), and therefore can be individually described by the Hamiltonian

$$\hat{H} = \beta_s \hat{a}_s^\dagger \hat{a}_s + \beta_f \hat{a}_f^\dagger \hat{a}_f, \tag{1}$$

with the bosonic annihilation (creation) operators $\hat{a}_{s/f}^{(\dagger)}$ for photons on the slow/fast optical axis with propagation constant $\beta_{s/f}$, respectively. As schematically shown in Fig. 2**A**, these axes may be oriented at an angle $\alpha$ towards the horizontal/vertical (H/V) frame of reference (*35*). Any value of $\alpha$ that deviates from integer multiples of $\pi/2$ then entails dynamics in the polarization states of photons propagating along the $z$-direction according to the Heisenberg equation of motion (see Supplementary Sec. 3 for details)

$$i\frac{d}{dz}\begin{pmatrix}\hat{a}_H^\dagger \\ \hat{a}_V^\dagger\end{pmatrix} = \begin{pmatrix}\bar{\beta} + \Delta \cdot \cos 2\alpha & \Delta \cdot \sin 2\alpha \\ \Delta \cdot \sin 2\alpha & \bar{\beta} - \Delta \cdot \cos 2\alpha\end{pmatrix}\begin{pmatrix}\hat{a}_H^\dagger \\ \hat{a}_V^\dagger\end{pmatrix}, \tag{2}$$

where $\bar{\beta} = (\beta_f + \beta_s)/2$ describes the mean propagation constant and $\Delta = (\beta_s - \beta_f)/2$ represents the strength of birefringence. Notably, this mathematical structure is fully equivalent to the description of the dynamics in a coupled and detuned two-waveguide system. In our polarization-coupling regime, the counterpart of the propagation constants' mismatch, $\Delta \cdot \cos 2\alpha$, and coupling strength, $\Delta \cdot \sin 2\alpha$, can both be adjusted via the birefringence $\Delta$ (*34*) and the rotation angle α (*35*). Here, the case of $\alpha = 45°$ – entailing equally large propagation constants for the H/V modes – appears particularly interesting since the system behaves analogously to a phase-matched directional coupler, or beam splitter, with a tunneling amplitude $t = \sin(z \cdot \Delta)$, depending on the propagation length $z$. Similar to conventional couplers (*36*), the performance of this component can be evaluated by measuring the quantum interference (*33*) between the polarization states. To this end, a polarization-duplexed input state $|H, V\rangle \equiv \hat{a}_H^\dagger \hat{a}_V^\dagger |0\rangle$ is synthesized from photon pairs generated by spontaneous parametric down-conversion and injected into a polarization-maintaining waveguide ($\alpha = 0$), in which a rotated section with angle $\alpha = (45.2 \pm 1.2)°$ and length $L$ is embedded (see Fig. 2B). As the time delay $\tau$ between the photons at the injection facet is varied, the coincidence counts of two avalanche photodiodes (APDs, registering output photons in H- and V-polarization, respectively) are recorded. The resulting two-dimensional "HOM landscape", obtained for 20 different lengths $L$, is shown in Figure 2**C**. Along the cross-section $\tau = 0$, the characteristic $\cos^2(2L \cdot \Delta)$ oscillation of the conventional two-mode coupler is reproduced with HOM interference visibilities (for fixed $L$) of up to $0.842 \pm 0.021$.

With the necessary tools at hand, we proceed to extend a system of two spatially coupled waveguides (evanescent coupling rate $\kappa$) to a square lattice encoded in space and polarization (Figs. 3**A,B**). Details on how birefringence establishes the second dimension of the graph structure are provided in the Supplementary (Sec. 4). By choosing $\kappa = \Delta$, the resulting square lattice constitutes a 2D hypercube (*32*) for which the unitary transfer matrix after a coupling length $L = \pi/(4\Delta)$ reads

$$U = \frac{1}{2}\begin{pmatrix}1 & i \\ i & 1\end{pmatrix}^{\otimes 2}, \qquad (3)$$

where $\otimes 2$ denotes the tensor square. In our experiment this is accomplished for a propagation length $L = 6.92$ cm. We investigate the collective dynamics of two-photon input states for all possible arrangements with at most one photon per site. After the transformation governed by the square lattice, the polarization components are separated by two on-chip polarization beam splitters (PBSs), with the photons subsequently detected by APDs. For distinguishable photons, the equally strong couplings between the lattice sites result in a uniform output probability distribution across the entire lattice (Fig. 3C). In contrast, as illustrated by the color-coded nonclassicality in Fig. 3**D**, for indistinguishable photons, destructive and constructive quantum interference causes full suppression ($v \approx -1$) and pronounced enhancement ($v > 0$), respectively. Upon closer examination, $v$ appears closely related to the self-inverse symmetries $S_1, S_2, S_3$ of the HC graph illustrated in Fig. 3**E**: In accordance with the HC-suppression law (*37–39*), the symmetry operation which leaves the input state invariant determines the suppression of four output states, while the detection probability of all other output states appears enhanced compared to distinguishable particles (cf. Fig. 3**F**).

In addition to increasing the dimensionality of lattices, Eq. (2) allows for even more general coupling scenarios: Combining waveguides with clockwise and counter-clockwise rotated fast axes (see Fig. 3**G**), the polarization coupling terms acquire opposite signs, thereby providing a direct route to implement negative coupling strengths without the need for complex background lattice engineering or ancillary waveguides (*40, 41*). Notably, an $\alpha = \pm 45°$ arrangement also enables coupling between the V-polarized mode of one waveguide to the H-polarized mode of the other, and vice versa (see Supplementary Sec. 6 for details). Following this approach, we implement a polarization-encoded square lattice in which one of the polarization couplings has a negative sign, thereby breaking the underlying symmetry of the graph with special implications: Individual photons cannot be found on the diagonally opposite site of their starting position as a consequence of destructive interference caused by the different signs of coupling strengths. The resulting output statistics for distinguishable and indistinguishable photons are shown in Figs. 3**H** and **I**, respectively, and clearly differ from those obtained with positive couplings only. Due to the absence of photons on specific sites, for each input arrangement the nonclassicality in Fig. 3I indicates suppression only for a single output state, whereas all other output states show no significant fingerprints of quantum interference ($v \approx 0$).

In a final set of experiments, we implement a three-dimensional (3D) quantum walk. An equilaterally coupled triangle of identical, birefringent waveguides (Fig. 4**A**), is transformed into a triangular prism whose unitary transfer matrix reads

$$U = \frac{1}{\sqrt{2}}\begin{pmatrix}1 & i \\ i & 1\end{pmatrix} \otimes A, \qquad (4)$$

where $A$ is the transfer matrix of the triangular subgraph (see Supplementary Sec. 4 for details). The input state $|H1, V1\rangle$ (one H- and one V-polarized photon injected into waveguide 1) is invariant under a HC symmetry, and, as shown in Fig. 4**B**, behaves similarly to the $S_1$-symmetric arrangement in the square lattice (cf. upper row in Fig. 3**F**): output states with one photon in each H and V polarization (shown in the central layer of the 3D two-photon graph in Fig. 4**B**) appear suppressed, such that both photons must exit the structure in the same polarization state. In contrast, the input states $|H1, H2\rangle$ and $|H1, V2\rangle$ violate HC symmetry, and, as a result, the influence of the hypercube structure vanishes. For instance, $|H1, H2\rangle$ partially suppresses the probability of finding both photons in different waveguides, and, at the same time, enhances bunching behavior of orthogonally polarized photons in the same waveguide (Fig. 4**C**). This property is inextricably linked to the bosonic QW on the triangular graph, independent from the polarization, since single photon hopping in the polarization degree of freedom in this case only accumulates irrelevant phases for the two-photon interference (see Supplementary Sec. 5 for details). Remarkably, the input state $|H1, V2\rangle$ leads to a purely fermionic anti-bunching behavior for output photons with different polarizations (details are provided in the Supplementary Sec. 5): Both photons, despite being bosons, inevitably end up in different waveguides, as evidenced by the full suppression and enhancement of the corner and remaining nodes in the central layer of the two-photon graph in Fig. 4**D**, respectively. The two paths, of both photons either maintaining or changing their polarization, resulting in different output polarizations, differ by a cumulative phase shift of $\pi$, such that anti-symmetric features of the two-particle wavefunction emerge (*42*). In other words, the two bosonic walkers behave as fermionic walkers on the equilateral triangular waveguide lattice and adhere to the degeneracy prohibition expressed by the Pauli principle.

The exploration of quantum dynamics on complex graphs plays an essential role in numerous disciplines of science. However, as the dimensionality increases, their experimental implementation becomes an ever more challenging task. We introduced a novel approach to expand the dimensionality of photonic lattices by utilizing the polarization degree of freedom as an additional synthetic dimension. As demonstrated by the observation of HOM interference in a single waveguide with appropriately tailored birefringence, the dynamical features introduced by the additional dimension are in perfect agreement with the characteristic quantum dynamics expected from spatial dimensions. In further proof-of-principle experiments, we observed quantum interference in fully controlled QWs of correlated photons on three dimensional graphs, a long-standing goal in quantum photonics. Beyond increasing the dimensionality of a given graph configuration, our technique naturally allows for the implementation of negative couplings without the need for any ancillary structures (*40, 41*). Along similar lines, polarization-coupled lattices can readily establish anti-bunching of photons – a key ingredient for probing complex fermionic QWs and harnessing their unique suppression behavior without relying on polarization-entangled two-photon sources (*43, 44*). As a direct result from the here established framework, a number of fascinating opportunities arise: The quantum dynamics of bilayer 2D materials can be emulated in photonic model systems, and the impact of nonlinearities on wave packets evolving on complex graphs becomes experimentally accessible by utilizing bright light pulses and the Kerr nonlinearity. Finally, topological questions such as the graph isomorphism problem can now be addressed more efficiently on optical platforms.

**References and Notes:**

1. M. Walschaers, J. F.-C. Diaz, R. Mulet, A. Buchleitner, Optimally Designed Quantum Transport across Disordered Networks. *Phys. Rev. Lett.* **111**, 180601 (2013).

2. S. N. Dorogovtsev, J. F. F. Mendes, Evolution of networks. *Adv. Phys.* **51**, 1079–1187 (2002).

3. R. Williams, The Control Of Neuron Number. *Annu. Rev. Neurosci.* **11**, 423–453 (1988).

4. J. Biamonte, M. Faccin, M. De Domenico, Complex networks from classical to quantum. *Commun. Phys.* **2** (2019), 53.

5. A. Acín, J. I. Cirac, M. Lewenstein, Entanglement percolation in quantum networks. *Nat. Phys.* **3**, 256–259 (2007).

6. G. D. Paparo, M. Müller, F. Comellas, M. A. Martin-Delgado, Quantum Google in a Complex Network. *Sci. Rep.* **3**, 2773 (2013).

7. B. Hein, G. Tanner, Quantum search algorithms on a regular lattice. *Phys. Rev. A*. **82**, 12326 (2010).

8. C. A. Ryan, M. Laforest, J. C. Boileau, R. Laflamme, Experimental implementation of a discrete-time quantum random walk on an NMR quantum-information processor. *Phys. Rev. A*. **72**, 062317 (2005).

9. H. Schmitz, R. Matjeschk, C. Schneider, J. Glueckert, M. Enderlein, T. Huber, T. Schaetz, Quantum walk of a trapped ion in phase space. *Phys. Rev. Lett.* **103**, 090504 (2009).

10. M. Karski, L. Förster, J. M. Choi, A. Steffen, W. Alt, D. Meschede, A. Widera, Quantum walk in position space with single optically trapped atoms. *Science (80-. ).* **325**, 174–177 (2009).

11. D. Bouwmeester, I. Marzoli, G. P. Karman, W. Schleich, J. P. Woerdman, Optical Galton board. *Phys. Rev. A - At. Mol. Opt. Phys.* **61**, 013410 (1999).

12. F. Zähringer, G. Kirchmair, R. Gerritsma, E. Solano, R. Blatt, C. F. Roos, Realization of a quantum walk with one and two trapped ions. *Phys. Rev. Lett.* **104**, 100503 (2010).

13. B. Do, M. L. Stohler, S. Balasubramanian, D. S. Elliott, C. Eash, E. Fischbach, M. A. Fischbach, A. Mills, B. Zwickl, Experimental realization of a quantum quincunx by use of linear optical elements. *J. Opt. Soc. Am. B*. **22**, 499–504 (2005).

14. H. B. Perets, Y. Lahini, F. Pozzi, M. Sorel, R. Morandotti, Y. Silberberg, Realization of quantum walks with negligible decoherence in waveguide lattices. *Phys. Rev. Lett.* **100**, 170506 (2008).

15. P. L. Knight, E. Roldán, J. E. Sipe, Quantum walk on the line as an interference phenomenon. *Phys. Rev. A*. **68**, 020301 (2003).

16. M. Walschaers, F. Schlawin, T. Wellens, A. Buchleitner, Quantum Transport on Disordered and Noisy Networks: An Interplay of Structural Complexity and Uncertainty. *Annu. Rev. Condens. Matter Phys.* **7**, 223–248 (2016).

17. K. Mayer, M. C. Tichy, F. Mintert, T. Konrad, A. Buchleitner, Counting statistics of many-particle quantum walks. *Phys. Rev. A*. **83**, 062307 (2011).

18. A. M. Childs, D. Gosset, Z. Webb, Universal computation by multiparticle quantum walk. *Science (80-. ).* **339**, 791–794 (2013).

19. J. K. Gamble, M. Friesen, D. Zhou, R. Joynt, S. N. Coppersmith, Two-particle quantum walks applied to the graph isomorphism problem. *Phys. Rev. A*. **81**, 052313 (2010).

20. A. Peruzzo, M. Lobino, J. C. F. Matthews, N. Matsuda, A. Politi, K. Poulios, X. Q. Zhou, Y. Lahini, N. Ismail, K. Wörhoff, Y. Bromberg, Y. Silberberg, M. G. Thompson, J. L. OBrien, Quantum walks of correlated photons. *Science (80-. ).* **329**, 1500–1503 (2010).

21. J. M. Harrison, J. P. Keating, J. M. Robbins, Quantum statistics on graphs. *Proc. R. Soc. A*. **467**, 212–233 (2010).

22. M. B. Plenio, S. F. Huelga, Dephasing-assisted transport: Quantum networks and biomolecules. *New J. Phys.* **10**, 113019 (2008).

23. M. Mohseni, P. Rebentrost, S. Lloyd, A. Aspuru-Guzik, Environment-assisted quantum walks in photosynthetic energy transfer. *J. Chem. Phys.* **129**, 174106 (2008).

24. D. N. Biggerstaff, R. Heilmann, A. A. Zecevik, M. Gräfe, M. A. Broome, A. Fedrizzi, S. Nolte, A. Szameit, A. G. White, I. Kassal, Enhancing coherent transport in a photonic network using controllable decoherence. *Nat. Commun.* **7**, 11282 (2016).

25. A. M. Childs, R. Cleve, E. Deotto, E. Farhi, S. Gutmann, D. A. Spielman, in *Conference Proceedings of the Annual ACM Symposium on Theory of Computing* (2003), pp. 59–68.

26. K. Poulios, R. Keil, D. Fry, J. D. A. Meinecke, J. C. F. Matthews, A. Politi, M. Lobino, M. Gräfe, M. Heinrich, S. Nolte, A. Szameit, J. L. O'Brien, Quantum walks of correlated photon pairs in two-dimensional waveguide arrays. *Phys. Rev. Lett.* **112**, 143604 (2014).

27. O. Boada, A. Celi, J. I. Latorre, M. Lewenstein, Quantum Simulation of an Extra Dimension. *Phys. Rev. Lett.* **108**, 133001 (2012).

28. L. Yuan, Q. Lin, M. Xiao, S. Fan, Synthetic dimension in photonics. *Optica*. **5**, 1396–1405 (2018).

29. E. Lustig, S. Weimann, Y. Plotnik, Y. Lumer, M. A. Bandres, A. Szameit, M. Segev, Photonic topological insulator in synthetic dimensions. *Nature*. **567** (2019), 356–360.

30. O. Zilberberg, S. Huang, J. Guglielmon, M. Wang, K. P. Chen, Y. E. Kraus, M. C. Rechtsman, Photonic topological boundary pumping as a probe of 4D quantum Hall physics. *Nature*. **553**, 59–62 (2018).

31. A. Szameit, S. Nolte, Discrete optics in femtosecond-laserwritten photonic structures. *J. Phys. B At. Mol. Opt. Phys.* **43**, 163001 (2010).

32. C. Dittel, R. Keil, G. Weihs, Many-body quantum interference on hypercubes. *Quantum Sci. Technol.* **2**, 015003 (2017).

33. C. K. Hong, Z. Y. Ou, L. Mandel, Measurement of subpicosecond time intervals between two photons by interference. *Phys. Rev. Lett.* **59**, 2044–2046 (1987).

34. P. Yang, G. R. Burns, J. Guo, T. S. Luk, G. A. Vawter, Femtosecond laser-pulse-induced birefringence in optically isotropic glass. *J. Appl. Phys.* **95**, 5280–5283 (2004).

35. C.-Y. Wang, J. Gao, X.-M. Jin, On-chip rotated polarization directional coupler fabricated

by femtosecond laser direct writing. *Opt. Lett.* **44**, 102–105 (2019).

36. G. D. Marshall, A. Politi, J. C. F. Matthews, P. Dekker, M. Ams, M. J. Withford, J. L. O'Brien, Laser written waveguide photonic quantum circuits. *Opt. Express*. **17**, 12546–12554 (2009).

37. N. Viggianiello, F. Flamini, L. Innocenti, D. Cozzolino, M. Bentivegna, N. Spagnolo, A. Crespi, D. J. Brod, E. F. Galvão, R. Osellame, F. Sciarrino, Experimental generalized quantum suppression law in Sylvester interferometers. *New J. Phys.* **20**, 033017 (2018).

38. C. Dittel, G. Dufour, M. Walschaers, G. Weihs, A. Buchleitner, R. Keil, Totally Destructive Many-Particle Interference. *Phys. Rev. Lett.* **120**, 240404 (2018).

39. C. Dittel, G. Dufour, M. Walschaers, G. Weihs, A. Buchleitner, R. Keil, Totally destructive interference for permutation-symmetric many-particle states. *Phys. Rev. A*. **97**, 062116 (2018).

40. J. M. Zeuner, M. C. Rechtsman, R. Keil, F. Dreisow, A. Tünnermann, S. Nolte, A. Szameit, Negative coupling between defects in waveguide arrays. *Opt. Lett.* **37**, 533 (2012).

41. R. Keil, C. Poli, M. Heinrich, J. Arkinstall, G. Weihs, H. Schomerus, A. Szameit, Universal Sign Control of Coupling in Tight-Binding Lattices. *Phys. Rev. Lett.* **116**, 213901 (2016).

42. J. M. Leinaas, J. Myrheim, On the theory of identical particles. *Nuovo Cim. B*. **37**, 1–23 (1977).

43. L. Sansoni, F. Sciarrino, G. Vallone, P. Mataloni, A. Crespi, R. Ramponi, R. Osellame, Two-particle bosonic-fermionic quantum walk via integrated photonics. *Phys. Rev. Lett.* **108**, 010502 (2012).

44. J. C. F. Matthews, K. Poulios, J. D. A. Meinecke, A. Politi, A. Peruzzo, N. Ismail, K. Wörhoff, M. G. Thompson, J. L. O'Brien, Observing fermionic statistics with photons in arbitrary processes. *Sci. Rep.* **3**, 1539 (2013).

45. L. A. Fernandes, J. R. Grenier, P. R. Herman, J. S. Aitchison, P. V. S. Marques, Femtosecond laser fabrication of birefringent directional couplers as polarization beam splitters in fused silica. *Opt. Express*. **19**, 11992–11999 (2011).

46. Y. Bromberg, Y. Lahini, R. Morandotti, Y. Silberberg, Quantum and classical correlations in waveguide lattices. *Phys. Rev. Lett.* **102**, 253904 (2009).

47. S. Rojas-Rojas, L. Morales-Inostroza, U. Naether, G. B. Xavier, S. Nolte, A. Szameit, R. A. Vicencio, G. Lima, A. Delgado, Analytical model for polarization-dependent light propagation in waveguide arrays and applications. *Phys. Rev. A*. **90**, 063823 (2014).

48. M. C. Tichy, M. Tiersch, F. Mintert, A. Buchleitner, Many-particle interference beyond many-boson and many-fermion statistics. *New J. Phys.* **14**, 093015 (2012).

49. J. L. Walsh, A Closed Set of Normal Orthogonal Functions. *Am. J. Math.* **45**, 5–24 (1923).

**Acknowledgments:** The authors would like to thank C. Otto for preparing the high-quality fused silica samples used for the inscription of all photonic structures employed in this work.

**Funding:** The authors acknowledge funding from the European Union Horizon 2020 program (ErBeStA 800942), the Deutsche Forschungsgemeinschaft (grants SZ 276/12-1, BL 574/13-1, SZ 276/15-1, SZ 276/20-1, SZ 276/21-1), the Austrian Science Fund (grant P 30459), the Georg H. Endress foundation, and the Alfried Krupp von Bohlen and Halbach foundation.

**Author contributions:** R.K. developed the theoretical framework of polarization coupling in waveguide circuits. M.E. performed the experiments and refined the theoretical model. M.E., M.H. and A.S. interpreted the measurement data. A.S. supervised the project. All authors discussed the results and co-wrote the paper.

**Competing interests:** The authors declare no competing interests.

**Data and materials availability:** All data is available in the main text or the supplementary materials.

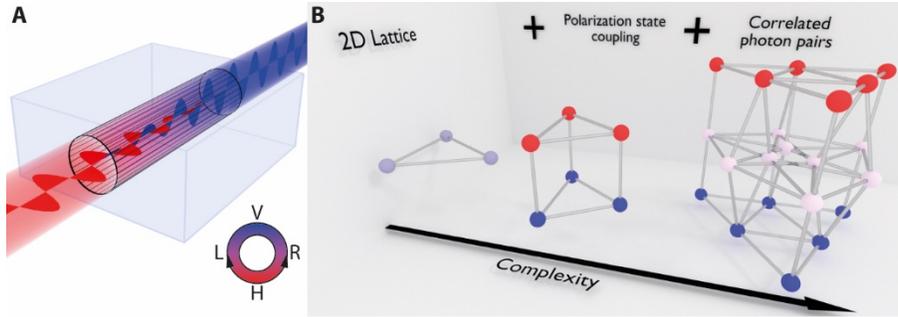

**Fig. 1.** Utilizing polarization as an additional synthetic dimension. (**A**) A single waveguide with tailored birefringence coherently couples its horizontally (red) and vertically (blue) polarized modes of the electromagnetic field. (**B**) Planar graphs (left) acquire an additional dimension due to the coupling of two polarization states (center). The Hilbert space of photon pairs on three-dimensional graphs takes the form of a yet more complex graph (right).

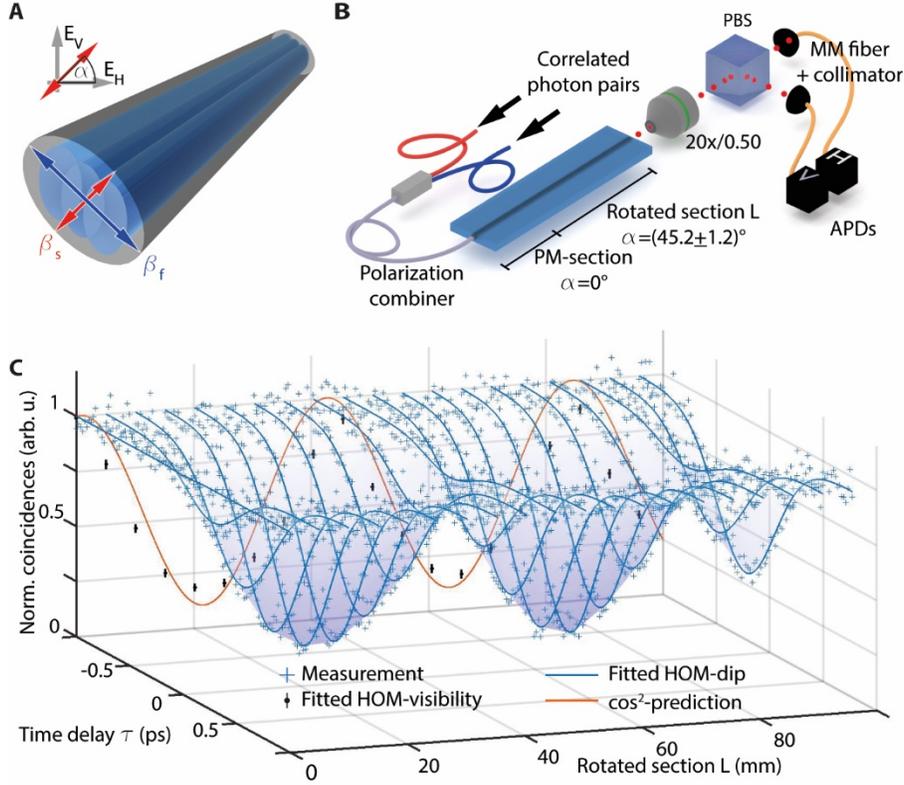

**Fig. 2.** Quantum interference in a polarization coupler. (**A**) Triple-pass-femtosecond-laser-written waveguides enable control over both magnitude and orientation of the birefringence. Changes to the angle α of the slow axis allow for polarization-maintaining (PM) sections to be included at will. (**B**) Correlated photon pairs combined in a single waveguide exhibit Hong-Ou-Mandel interference due to a coupling of the horizontal and vertical polarization modes in a section with rotated fast and slow axes of length *L*. (**C**) Coincidence rate measured as a function of the time delay $\tau$ between the photons' arrival time and length *L* of the rotated section. The displayed $\cos^2$-prediction fits the data for $\tau = 0$ and a visibility limited only by the photon source to (92.3±1.1)% (see Supplementary Sec. 1 for details). The largest observed visibility was (84.2 ± 2.1)%.

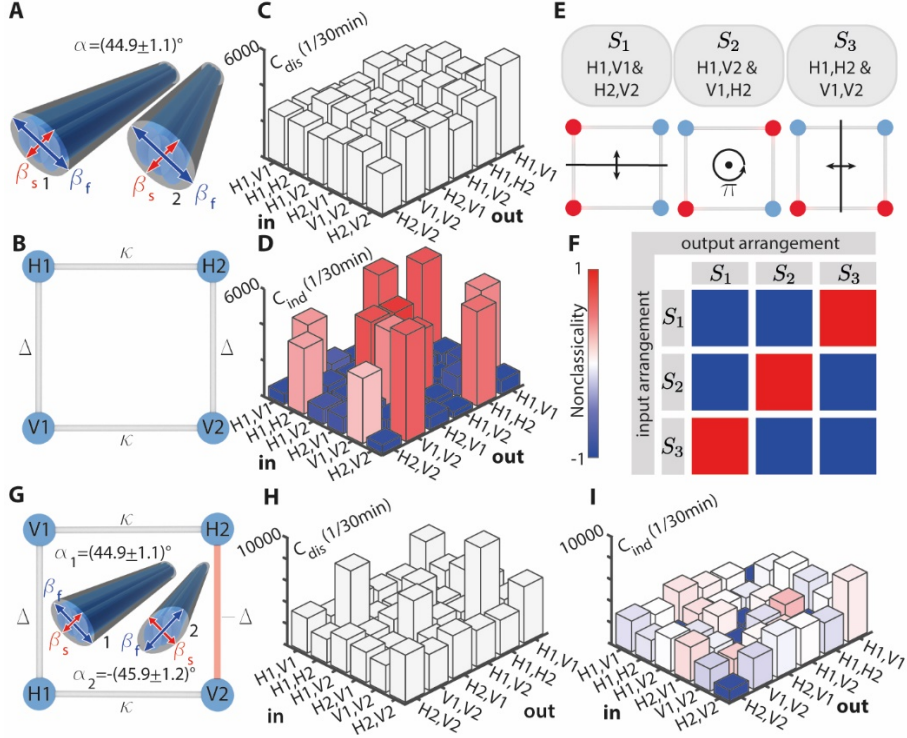

**Fig. 3.** Square lattices in a single spatial dimension. (**A**) A pair of birefringent waveguides with slow axes at $\alpha = 45°$ implements a square graph structure (**B**) with coupling strengths $\kappa = (\kappa_f + \kappa_s)/2$ (averaged coupling strength between slow/fast axes) between the waveguides 1 and 2, and $\Delta = (\beta_s - \beta_f)/2$ between the polarizations H and V. (**C**) Measured coincidence matrix of two distinguishable photons. For coupling parameters $\kappa = \Delta$, classical (single-particle) dynamics yields a nearly uniform distribution after a propagation length of $L = \pi/(4\Delta)$. (**D**) Coincidence rates for indistinguishable photons. In addition to the count rates (height), the nonclassicality $\nu$ is color-coded to indicate enhancement (red) and suppression (blue) by quantum (two-particle) interference. (**E**) Illustration of the self-inverse symmetry operations $S_1, S_2, S_3$ on the 2D hypercube graph, leaving the two-photon arrangements (indicated by red spheres) invariant. (**F**) Nonclassicality for combinations of initial-final photon arrangements captured by the suppression laws associated with $S_1$, $S_2$, and $S_3$. The red (blue) squares indicate enhanced (reduced) probabilities of the corresponding input-output transition. (**G**) Clockwise ($\alpha = +45°$) and counter-clockwise ($\alpha = -45°$) rotated optical axes implement a square lattice with negative coupling strength between two sites. For equal coupling strengths $\kappa = \Delta$ ($\kappa$: coupling strength between fast and slow axes of different waveguides) and propagation length $L = \pi/(4\Delta)$, we measure the coincidence rates for (**H**) distinguishable and (**I**) indistinguishable photon pairs.

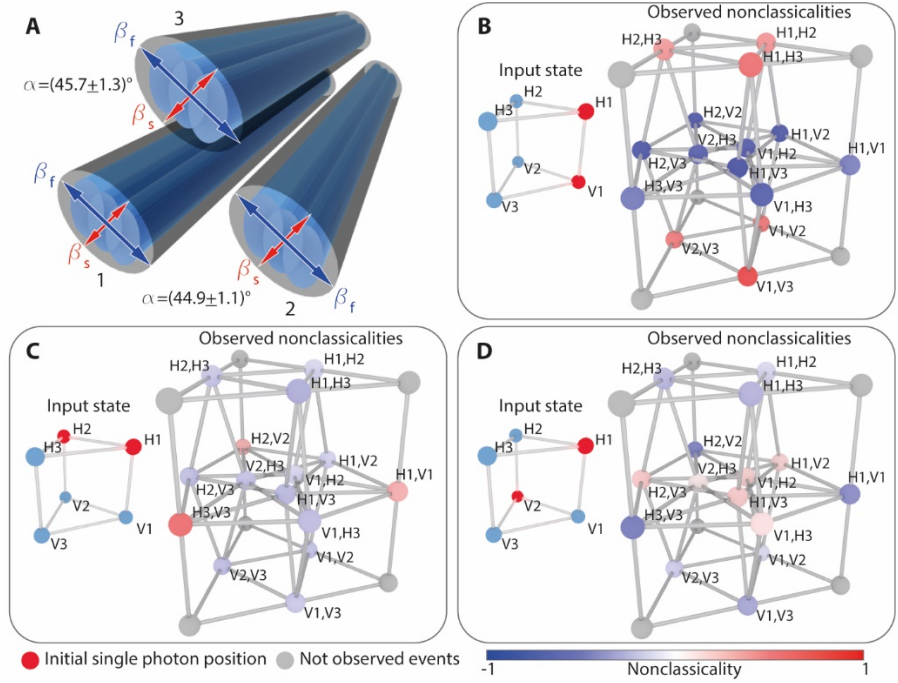

**Fig. 4.** Three-dimensional graph in two spatial dimensions. (**A**) The graph structure of a triangular prism is realized with three coupled birefringent waveguides arranged in the shape of an equilateral triangle. (**B**-**D**) Two-photon input states are illustrated by red nodes on the single photon graphs, and the corresponding experimentally observed nonclassicalities (coincidence rates are available in the Supplementary Fig. S4) are color-coded on a two-photon graph representation. Grey nodes indicate output states with both photons in the same mode and polarization, which are inaccessible in the present experimental setting without photon-number-resolving detection.